# Magnetorotational Mechanism of the Explosion of Core-Collapse Supernovae

G. S. Bisnovatyi-Kogan[1,2]*, S. G. Moiseenko[1]**, and N. V. Ardelyan[3]***

[1] Space Research Institute, Russian Academy of Sciences, Profsoyuznaya ul. 84/32, Moscow, 117997 Russua.
[2] National Research Nuclear University MEPhI, Kashirskoe sh. 31, Moscow, 115409 Russia.
[3] Faculty of Computational Mathematics and Cybernetics, Moscow State University, Moscow, 119991 Russia.
*E-mail: gkogan@iki.rssi.ru
**E-mail: moiseenko@iki.rssi.ru
***E-mail: ardel@cs.msu.su

**Abstract**—The idea of the magnetorotational explosion mechanism is that the energy of rotation of the neutron star formed in the course of a collapse is transformed into the energy of an expanding shock wave by means of a magnetic field. In the two-dimensional case, the time of this transformation depends weakly on the initial strength of the poloidal magnetic field because of the development of a magnetorotational instability. Differential rotation leads to the twisting and growth of the toroidal magnetic-field component, which becomes much stronger than the poloidal component. As a result, the development of the instability and an exponential growth of all field components occur. The explosion topology depends on the structure of the magnetic field. In the case where the initial configuration of the magnetic field is close to a dipole configuration, the ejection of matter has a jet character, whereas, in the case of a quadrupole configuration, there arises an equatorial ejection. In either case, the energy release is sufficient for explaining the observed average energy of supernova explosion. Neutrinos are emitted as the collapse and the formation of a rapidly rotating neutron star proceeds. In addition, neutrino radiation arises in the process of magnetorotational explosion owing to additional rotational-energy losses. If the mass of a newborn neutron star exceeds the mass limit for a nonrotating neutron star, then subsequent gradual energy losses may later lead to the formation of a black hole. In that case, the energy carried away by a repeated flash of neutrino radiation increases substantially. In order to explain an interval of 4.5 hours between the two observed neutrino signals from SN 1987A, it is necessary to assume a weakening of the magnetorotational instability and a small initial magnetic field ($10^9$–$10^{10}$ G) in the newly formed rotating neutron star. The existence of a black hole in the SN 1987A remnant could explain the absence of any visible pointlike source at the center of the explosion.

## 1. INTRODUCTION

Supernovae are among the most powerful explosions in the Universe. Their average observed energy in the form of kinetic energy and energy of radiation is about $10^{51}$ erg. Observations of supernova bursts made it possible to obtain deeper insight into the current properties of the Universe, and the supernovae themselves are the locus of production of new objects that have extreme properties. Observations of exploding type-Ia supernovae were of crucial importance for the discovery of dark energy; neutron stars and black holes arise upon the explosions of core-collapse supernovae; supernovae and their remnants are the locus of production of high-energy cosmic rays; etc.

Supernovae arise at the end of the evolution of rather massive stars in the main sequence, $M_{ms} >= 8\ M_{sun}$. A schematic picture of the evolution of stars with various masses from the main sequence to the formation carbon–oxygen white dwarfs is shown in Fig. 1, which was borrowed from [1]. The evolution of a star with initial mass $25 M_{sun}$ ends up in core collapse and the burst of a type-II supernova.

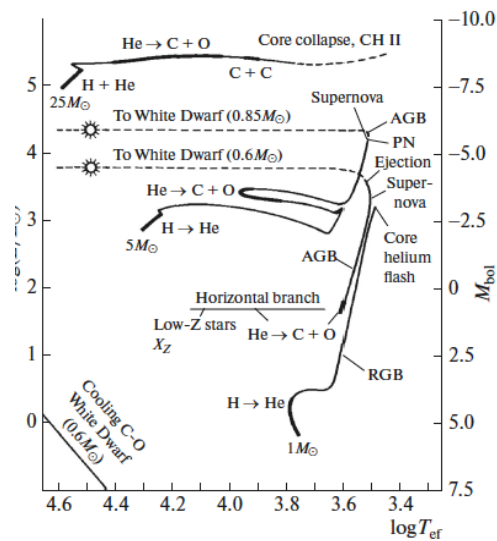

**Fig. 1.** Evolutionary tracks in the Hertzsprung–Russell diagram for various star masses from the main sequence to the end of evolution [1].

A supernova may explode at the end of two evolutionary trajectories. As the result of evolution, stars of medium mass (about 8 to 12 M_) develop a carbon–oxygen core of mass exceeding the Chandrasekhar limit of about 1.4 $M_{sun}$. Therefore, this core loses stability and begins undergoing contraction. The growth of temperature leads to the thermonuclear explosion of the carbon–oxygen core and to its complete disruption observed in the form of a type-Ia supernova. At an initial mass not lower than 12 $M_{sun}$, the degree of core degeneracy is so low that a quiet evolution may proceed up to the formation of an iron core, which loses stability, collapsing into a neutron star. This leads to the observed burst of supernovae belonging to SN II and SN Ib types and their various subtypes. The formation of a neutron star releases an enormous energy of about $5 \times 10^{53}$ erg, which is equal to the gravitational binding energy of a neutron star. The core-collapse supernova mechanism involving the formation of a neutron star was first proposed in the pioneering study of Baade and Zwicky [2]. Exploding supernovae were considered by Hoyle and Fowler [3]. The first calculations devoted to collapse followed by the formation of a neutron star and to the subsequent behavior of the envelope were performed by Colgate and White [4]; Arnett [5, 6]; and Ivanova, Imshennik, and Nadyozhin [7]. Those calculations were performed for nonrotating stars in the spherically symmetric approximation. It was assumed that the flow of neutrinos from the collapsing core could lead to surrounding-envelope heating (neutrino deposition) sufficient for shock formation, envelope ejection, and the appearance of a supernova. More precise calculations performed by Nadyozhin [8–10] revealed an insufficient efficiency of neutrino deposition and the absence of an explosion in the one-dimensional spherically symmetric model. Further investigations into the neutrino model of the explosion of core-collapse supernovae involved taking into account deviations from spherical symmetry. In 1979, Epstein [11] showed that a superadiabatic temperature gradient arises in the outer layers of a nascent neutron star and that the a convective instability develops there. Convection carries outside hotter matter from inner star layers, thereby increasing the energy of each emitted neutrino. By virtue of the growth of the weak-interaction cross section with energy, more energetic neutrinos heat envelope matter more strongly, which could lead to the burst of a supernova. The calculations revealed that this is not so. The advent of ever faster computers made it possible to perform two- and three-dimensional calculations of collapse with allowance for rotation. In doing this, there was a hope for the transport of higher energy neutrinos from deeper layers and for the attainment of an explosion. Various two- and threedimensional calculations (for an overview, see [12, 13]) did not lead to an unambiguous conclusion. In three-dimensional calculations, it is more difficult to obtain an explosion than in two-dimensional calculations, since, in a realistic three-dimensional situation, the fragmentation of convective vortices occurs, so that their size becomes smaller, in contrast to what we have in the two-dimensional model, where there is the merger of vortices, so that their size grows with time.

## 2. MAGNETOROTATIONAL MODEL OF EXPLOSION: ONE-DIMENSIONAL CALCULATIONS

After the discovery of radio pulsars (see the article of Hewish and his coauthors that was published in 1968 [14]), it became clear that neutron stars rotated quickly and featured a strong magnetic field. In this connection, there appeared the idea that the rotational energy developed by a neutron star originating from collapse and used via the magnetic field is the source of supernova-explosion energy. A qualitative model of magnetorotational explosion was proposed in [15]. This explosion model was supported by the following observational facts. Almost all remnants of supernova explosions have a shape far from a spherically symmetric shape. All stars rotate and have magnetic fields that, upon collapse, in which the magnetic flux is conserved, should grow up to strengths observed in radio pulsars [16]. Young remnants of core collapse supernovae, such as Crab and Vela (see [17, 18]), feature directed, possibly one-sided, ejections (jets). These properties cannot be explained within a simple spherically symmetric explosion model.

Qualitatively, the magnetorotational-explosion picture employing a magnetic field as a transition belt for transforming the rotational energy into the explosion energy looks as follows [15]. The neutron star resulting from collapse rotates nonuniformly in such a way that the angular velocity of rotation decreases from the center to the periphery. Under conditions of a nonuniform rotation, the magnetic field force lines undergo twisting, with the result that the field grows fast. An increase in the magnetic-field pressure, together with angular-momentum flux from the center to the periphery, leads to the appearance of perturbations, which, propagating in the neutron star through a medium characterized by a decreasing density, grow. Owing to this, the magnetosonic wave transforms into a shock wave, whereupon the ejection of the envelope and an explosion occur.

The first one-dimensional calculations for a magnetorotational explosion [19] revealed a high efficiency (about 10%) of rotational-energy transformation into the explosion energy by means of the magnetic field. For a neutron star of mass 1.4 $M_{sun}$, the ejected mass was about 0.1 $M_{sun}$, while its kinetic energy, which was nearly equal to the explosion energy, was about $10^{51}$ erg. The calculation was performed within a cylindrical model where the neutron star was represented as a cylindrical plate (see Fig. 2).

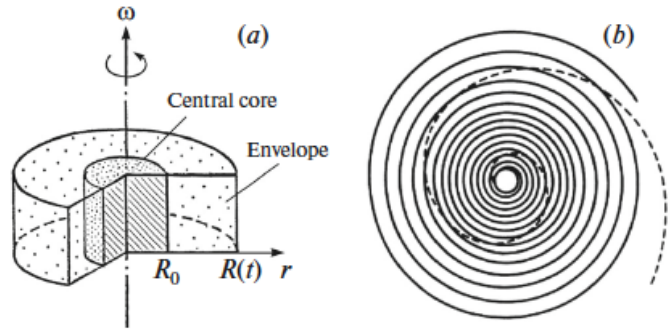

**Fig. 2.** (*a*) Schematic picture of the initial state of the model as a unit-length cylinder [19] and (*b*) configuration of magnetic-field lines of force at the instant of $t = 7/\sqrt{\alpha}$ for (dashed curve) $\alpha = 0.01$ and (solid curve) $\alpha = 10^{-4}$ in the vicinity of the core rotating as a rigid body [20].

The equations of ideal magnetic hydrodynamics were solved in the approximation of cylindrical symmetry. The core was assumed to rotate as a rigid body, and the solution was sought only in the envelope. A differential rotation and a radial magnetic field were present at the initial instant in the envelope. The most important special features of superdense neutron-star matter were taken into account approximately in the equation of state. They include the degeneracy of electrons and a relativistic character of their motion, nuclear interaction of nucleons, and temperature corrections. Neutrino-induced cooling via Urca processes was also taken into account by employing approximate formulas from [7]. The continuity of the angular velocity at the core–envelope interface was assumed, in which case the velocity of rigid-body core rotation was equal to the envelope velocity at the inner boundary. The latter in turn was determined from the condition requiring total-angular-momentum conservation for the core plus envelope system. In the case of preset velocity and radial dependence of the angular velocity, the initial ratio of the magnetic-field energy to the rotational energy of the system, α, is the only parameter in the problem being considered:

$$\alpha = \frac{E_{mag0}}{E_{rot0}}. \qquad (1)$$

The calculations reveal that the ejected mass and the explosion energy are virtually independent of the parameter α and that the characteristic time from the collapse to the beginning of the explosion, $t_{expl}$, is in inverse proportion to the initial strength of the magnetic field; that is,

$$t_{expl} \sim \frac{1}{\sqrt{\alpha}}. \qquad (2)$$

A dependence of this type stems from the fact that the explosion occurs when the magnetic pressure of the toroidal field comes to be on the same order of magnitude as the pressure of matter. In view of a linear growth of the field strength with time, $B_\varphi \approx B_{ini}$ ($t/t_0$), the respective time interval is inversely proportional to the strength of the initial radial field, $B_{ini}$. A strongly twisted toroidal field is shown in Fig. 2 at a small value of α. At small values of α, the time $t_{expl}$ is much longer than the explosion time, which is determined by the characteristic hydrodynamic time. In the case of employing explicit numerical schemes, the time step is bounded, according to the Courant condition, by the hydrodynamic time. This would lead to very long-term computations because of an increase in the number of time steps, possibly entailing the loss of accuracy via the accumulation of numerical errors. Therefore, implicit finite-difference schemes were applied in numerically solving "hard" systems of equations [21]. In our two-dimensional calculations, an implicit fully conservative Lagrangian finite-difference scheme on a remapping triangular grid was used [22–24]. The initial arrangement of triangular cells is shown in Fig. 3 for the whole star (left-hand panel) and for its central part (right-hand panel). At the present time, several international research groups are involved in the simulation of the magnetorotational mechanism of the explosion of core collapse supernovae (see, for example, [25–27]).

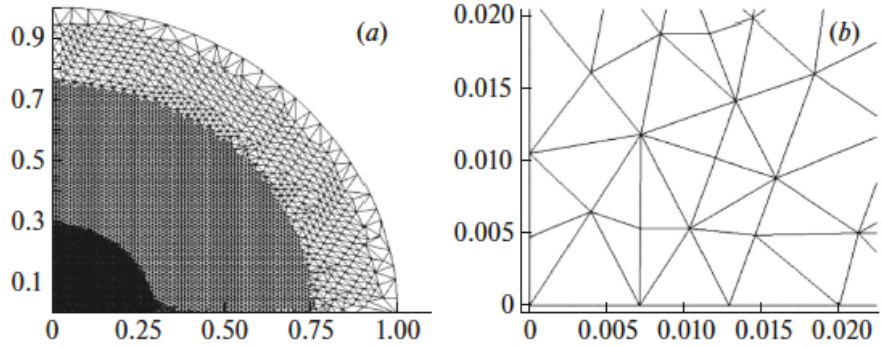

Fig. 3. Initial arrangement of the triangular grid cells for (*a*) the whole star and (*b*) its central part.

### 3. MAGNETOROTATIONAL SUPERNOVA: TWO-DIMENSIONAL MODEL

The first step in the two-dimensional approximation with allowance for self-gravitation under the assumption of axial symmetry ( $\partial/\partial\varphi = 0$) and equatorial symmetry with respect to the z = 0 plane involved calculating the collapse of a rotating iron white dwarf with mass 1.2Msun from an initial static unstable state to the formation of a steady-state neutron star [28]. As the result of collapse, a neutron star executing a strong differential rotation, whose angular-velocity profile is shown in Fig. 4, was formed from the white dwarf rotating as a rigid body. The calculations of the collapse were performed without allowance for the magnetic field, which, in view of its relative smallness, did not affect the contraction process.

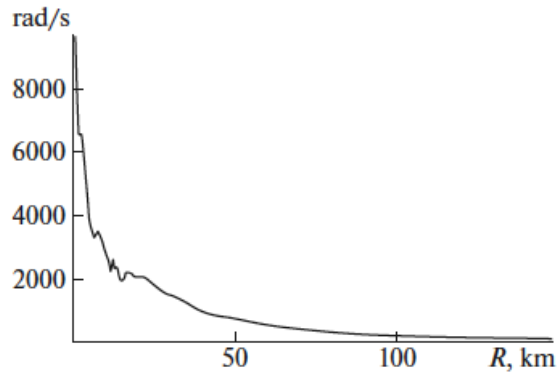

Fig. 4. Angular-velocity profile for a stationary rotation of the neutron star upon the collapse of an iron white dwarf having the mass 1.2 $M_\odot$ and rotating as a rigid body.

A magnetic field was included in the stationary model of a rotating neutron star without changing substantially the model as such, but, because of the twisting of the magnetic-field lines of force, its stationarity was violated. A strong increase in the magnetic field led to the emergence of a shock wave that caused the ejection of the envelope. Two-dimensional calculations within the magnetorotational supernova model were first performed for a quadrupole-like initial magnetic field [29]. In the course of the calculations, the triangular grid in Fig. 3 underwent an automatic remapping such that the grid density increased in central part of the core and decreased at the periphery, nearly without a change in the total number of cells (about 20 000). Remapping made it possible to reach a satisfactory accuracy of the calculations for a relatively small number of cells. For an initial magnetic field of quadrupole topology, the ejection of the envelope proceeded basically around the equatorial plane, since the radial component of the twisted field had a maximum strength at the equator. In the calculations for the model where the initial magnetic field had a dipole character [30], the ejected mass had nearly the same value as in the quadrupole model. The time dependences of the ejected mass and the ejected-envelope energy for the quadrupole model are presented in Fig. 5.

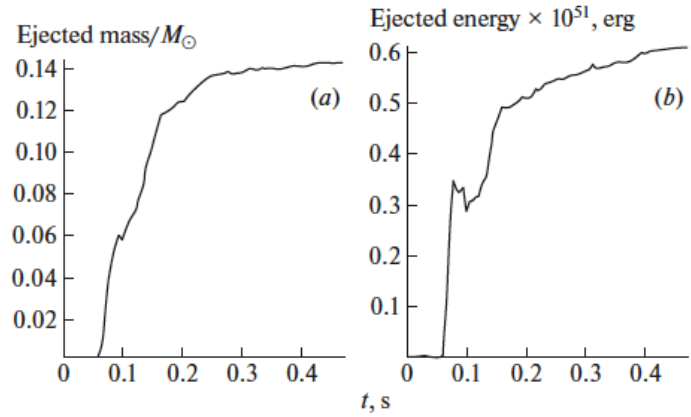

**Fig. 5.** Time dependence of the (*a*) ejected mass (in $M_\odot$ units) and (*b*) ejected-envelope energy (in erg units) in the course of a magnetorotational explosion for the quadrupole configuration of the initial magnetic field (matter is taken to be ejected if its kinetic energy is greater than the gravitational energy and if its velocity has an outward direction, $\alpha = 10^{-6}$).

A significant distinction between the quadrupole and dipole models was observed in the topology of ejections. In the case of a dipole initial configuration, the ejection of matter proceeded in the form of a weakly collimated jet along the axis of rotation, since regions where the radial field components had maximum strengths lay near this axis. Observations show the presence of directed ejections that were formed upon supernova explosions and which are seen in the photographs of young supernova remnants—Crab Nebula from [31] and the remnant in the Vela constellation from [32] in Fig. 6, as well as the Cassiopeia A supernova remnant [33]—and in the observations [34] of SN 2006gy (Fig. 7).

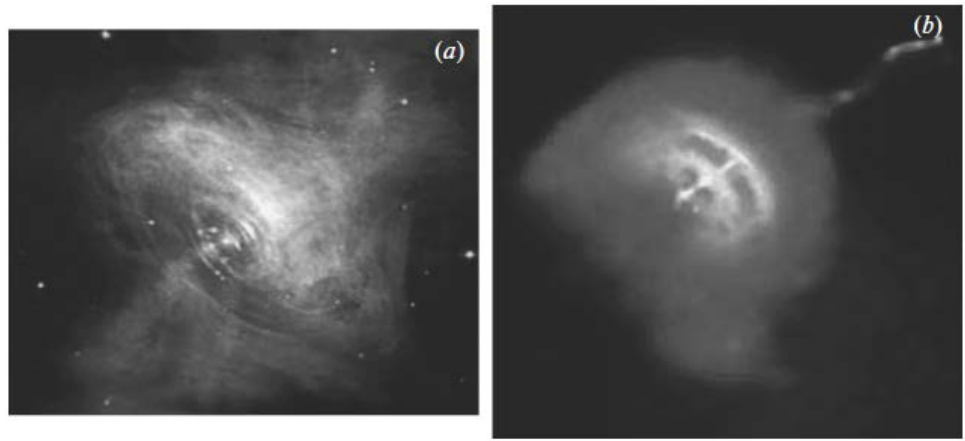

**Fig. 6.** (*a*) Joint optical–x-ray image of the Crab Nebula with the synchrotron radiation of the nebula formed by the pulsar wind resulting from the escape of particles and the magnetic field from the central pulsar. (*b*) Chandra x-ray image of the radio pulsar in the remnants of the Vela supernova with its pulsar-wind nebula.

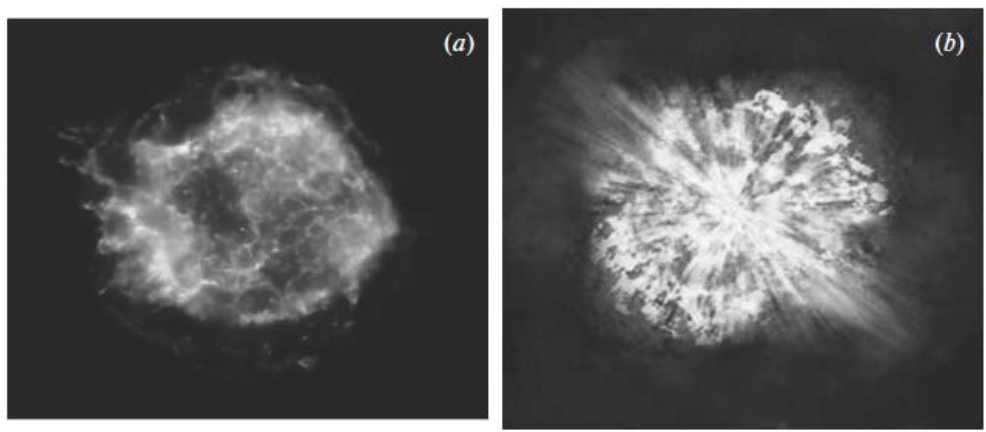

**Fig. 7.** (*a*) Chandra x-ray image of the Cassiopeia A supernova remnant. (*b*) NASA's artist picture of SN 2006gy, which is one of the brightest hypernovae observed thus far.

The origin of ejections observed in the supernova remnants in Crab and Vela is usually attributed to the action of young pulsars. However, the origin of the ejection from Cassiopeia A, where no pulsar was observed, as well as the hypothesized picture of SN 2006gy explosion, may be associated with directed ejections upon a magnetorotational explosion in the case of a dipole field configuration. Calculations of magnetorotational explosion for various values of the core mass and the initial energy of rotation were performed in [35, 36], and their results are presented in Fig. 8. Obviously, the explosion energy grows substantially with increasing core mass and initial energy of rotation.

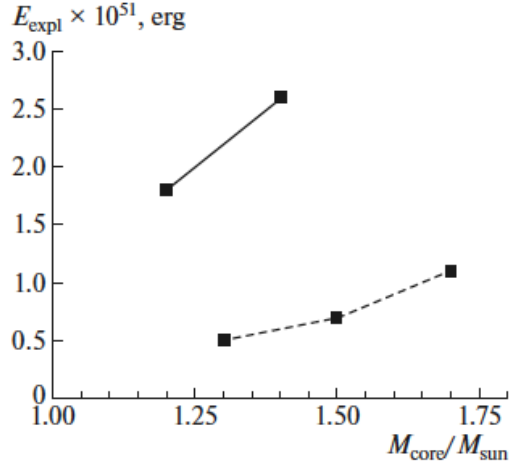

**Fig. 8.** Energy of the magnetorotational explosion of a supernova as a function of the initial core mass for the following values of the specific energy of matter rotation in the core: (solid curve) $E_{rot}/M_{core} = (0.39-0.40) \times 10^{19}$ erg/g and (dashed curve) $E_{rot}/M_{core} = (0.19-0.23) \times 10^{19}$ erg/g [35].

Figure 9 shows the quadrupole and dipole configurations of the initial magnetic field, while Fig. 10 gives the distribution of the toroidal magnetic field at the instant when the magnetic energy reaches a maximum value in the process of explosion for a quadrupole field configuration. The maximum strength of the magnetic field within a neutron star in the course of explosion is $H_{max} = 2.5 \times 10^{16}$ G. After the explosion, the magnetic-field strength at the star surface is $H \approx 4 \times 10^{12}$ G, which, under the condition of magnetic-flux conservation, corresponds to the field strength upon collapse. In the vicinity of the neutron-star surface, the magnetic field has a chaotic character and, locally, reaches a strength of $H \approx 2.5 \times 10^{14}$ G.

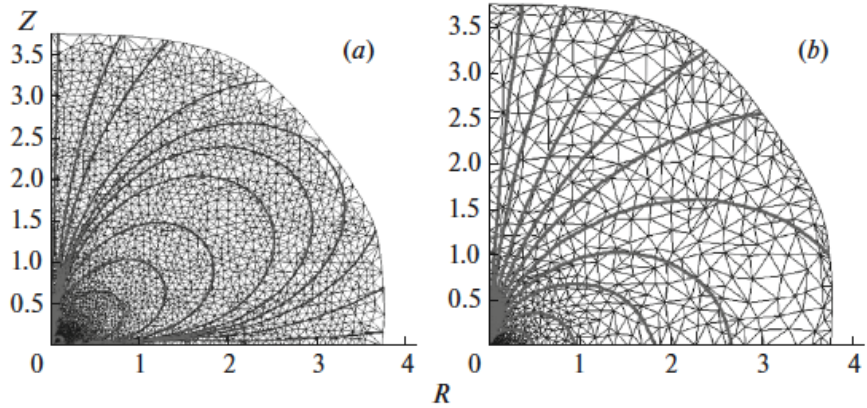

**Fig. 9.** (*a*) Quadrupole and (*b*) dipole initial field configurations used in the calculations reported in [29, 30].

## 4. MAGNETO-DIFFERENTIAL-ROTATIONAL INSTABILITY UPON EXPLOSION

In calculating magnetorotational explosion for various values of α from (1), it was noticed that, as α decreases, the time to the beginning of the explosion, $t_{expl}$, grows much more slowly than according to expression (2) in the one-dimensional calculations. Instead of being in inverse proportion to the strength of the initial magnetic field, the time $t_{expl}$ showed only a logarithmic growth as α decreases—specifically, it was $t_{expl} \approx 0.13$ s at $\alpha = 10^{-6}$ and $t_{expl} = 0.4$ s at $\alpha = 10^{-12}$, so that

$$t_{expl} \sim -\log \alpha \qquad (3)$$

In one-dimensional calculations, a decrease in α by six orders of magnitude would lead to the growth of $t_{expl}$ by a factor of $10^3$. An analysis revealed that a fast growth of the field upon the magnetorotational explosion and a slow growth of $t_{expl}$ were due to the development of the magnetorotational instability in a special form intimately related to the differential rotation of neutron-star matter. We call it a magneto–differential–rotation instability (MDRI). A simplified model describing the development of MDRI and leading to a logarithmic growth of $t_{expl}$ (see Fig. 11) was constructed in [29, 30].

Conditions for the development of MDRI differ from the conditions of magnetorotational instability (MRI) discovered in plasma [37, 38] and frequently used at the present time to explain the development of turbulence in accretion disks in x-ray binaries [39]. In accretion disks, MRI develops in the presence of a poloidal field that exceeds substantially the toroidal field and which is orthogonal to the plane of the rotating accretion disk.

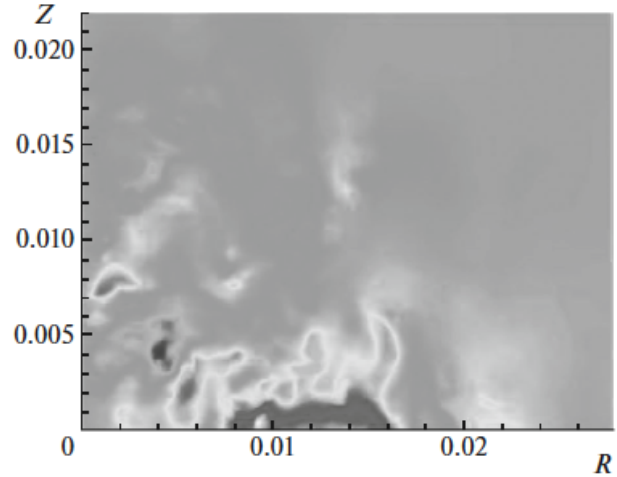

Fig. 10. Distribution of the toroidal magnetic field at the instant of its maximum magnetic energy in the course of explosion for the quadrupole initial field configuration [29].

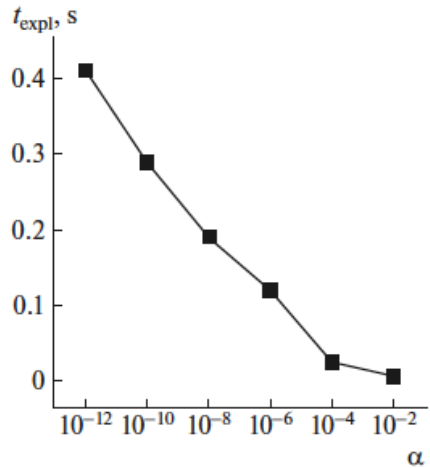

Fig. 11. Nearly logarithmic (because of the development of MDRI) of the time from the field-twisting instant to the explosion, $t_{expl}$, on the parameter $\alpha$ for a dipole-type magnetic field [30].

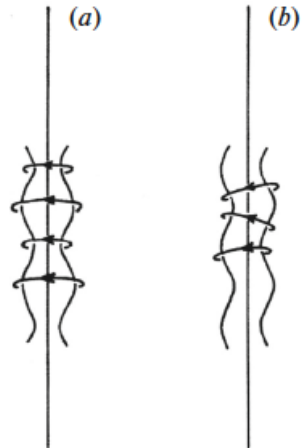

Fig. 12. Development of a magnetohydrodynamic instability in (*a*) two and (*b*) three dimensions within a non-rotating cylinder in the presence of a toroidal magnetic field [40]. The development of this instability occurs upon the growth of the toroidal component of the magnetic field because of differential rotation.

In a magnetorotational supernova, the development of MDRI occurs in response to the growth of the toroidal field in a differentially rotating star as soon as the toroidal component becomes substantially greater than the poloidal component. An instability of this type for a nonrotating cylinder was studied in [40] (see Fig. 12).

The development of MDRI was considered in [41–43]. An analysis of the calculations performed in [29, 30] made it possible to single out stages in the development of this instability that include the initial growth of the toroidal field owing to the twisting of the radial field component under conditions of differential rotation, the development of perturbations in the form of convective vortices in the meridional plane of the star being considered, and an enhancement of the radial field component by the differential rotation of these convective vortices and the subsequent increase in the rate of growth of the toroidal field. Thus, there arises a system featuring a positive feedback, where the field grows exponentially (see Fig. 13).

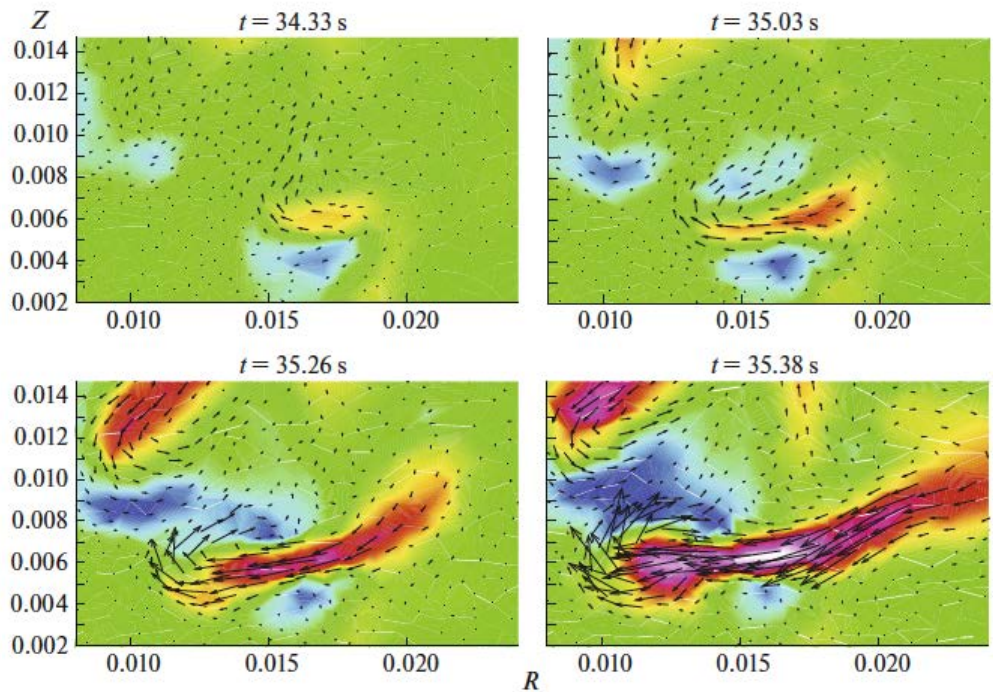

Fig. 13. Development of MDRI in the inner regions of differentially rotating stars. The growth of the toroidal field is shown as contour plots. Its strength corresponds to the length of the arrows [43].

The collapse of a rotating star in a uniform magnetic field was calculated by employing an explicit method of computations on an Eulerian grid [44]. The initial magnetic fields were chosen to be very large. At the presupernova magnetic field of strength $10^{12}$G, a bounce-off occurred owing to a strong magnetic field, and collapse gave way to disruption. In that case, MDRI did not develop. At the initial field strength of $10^9$ G, which was still much greater that the real one, matter returned back after the first bounce-off, whereupon the ejection of matter occurred owing to a magnetorotational explosion. Here, an equation of state that took into account basic properties of superdense matter was employed [45]. Similar calculations by means of a Lagrangian implicit method on a remapping triangular grid that were performed in [41–43] showed qualitative agreement of the results. Equations of state of various degrees of accuracy were used, but the difference in the respective results was found to be small. Qualitative distinctions were found upon a change in the magnetic field. Basically, they confirmed the results obtained in [44]. The time dependence of various forms of energy in the course of collapse for the initial magnetic field of strength $10_9$ G is shown in Fig. 14 from [41], along with the picture of fully developed MDRI.

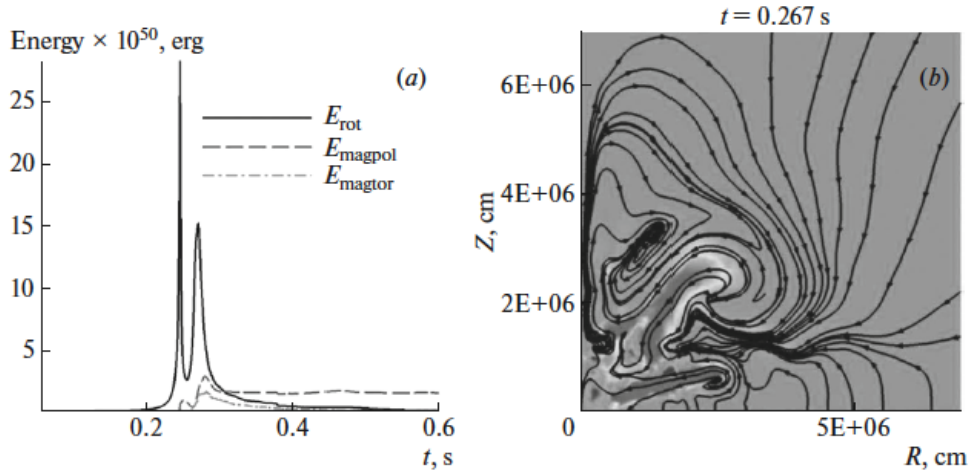

**Fig. 14.** (*a*) Time evolution of the (solid curve) rotational energy $E_{rot}$, (dashed curve) magnetic poloidal energy $E_{magpol}$, and (dash-dotted curve) magnetic toroidal energy $E_{magtor}$ for the initial field of strength $10^9$ G before the collapse and $E_{rot0}/E_{grav0} = 0.01$. (*b*) Fully developed stage of MDRI at the instant of $t = 0.267$ s. The toroidal magnetic field is shown by contour plots, and the lines of force of the poloidal magnetic field are represented by black curves with arrows [41, 43].

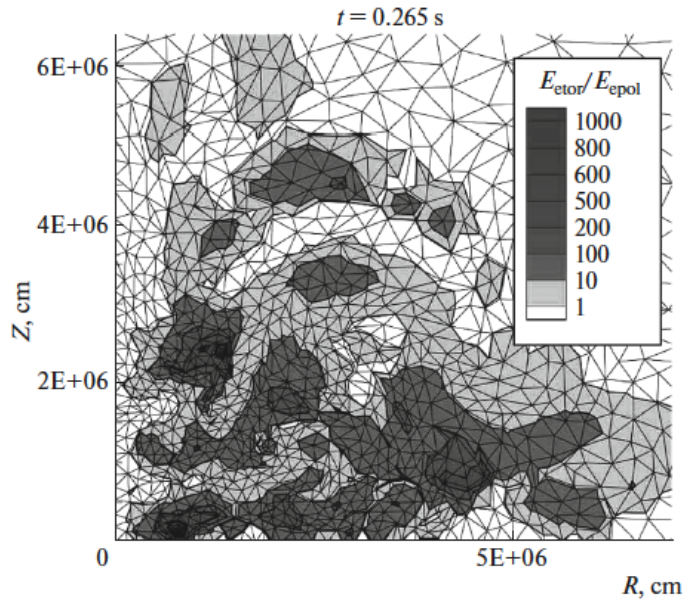

**Fig. 15.** Lagrangian triangular grid and ratio of the toroidal magnetic energy $E_{etor}$ to the poloidal magnetic energy $E_{epol}$ for each cell, $\frac{E_{etor}}{E_{epol}}$, at the instant of $t = 265$ ms for a uniform initial field of strength $H_0 = 10^9$ G and $E_{rot0}/E_{grav0} = 0.01$ [43].

Figure 15 from [43] gives the distribution of regions where there is a strong excess of the toroidal magnetic energy at the nonlinear stage of the process, which are seeds for the development of the instability.

## 5. ASYMMETRY OF EJECTIONS AND ORIGIN OF RAPIDLY MOVING PULSARS

The first observations indicating that the spatial vector of the velocity of motion of a pulsar is nearly aligned with its angular-velocity vector were presented in 1996 [46]. An observational confirmation of this conclusion was obtained in studies performed nine to sixteen years later. In [47], this conclusion was drawn from polarization observations of 25 pulsars. Linearly polarized radiation whose polarization vector is directed nearly along the magnetic-field axis was observed in 20 of them. This led the authors of [47] to the conclusion that the vector of the velocity of motion of a neutron star at the instant of its formation was aligned with the axis of its rotation. Further observations were aimed at excluding the influence of selection effects on this conclusion [48, 49]. As a result, it was shown that this correlation is confirmed by further polarization observations and that the aforementioned axes are aligned with an uncertainty not exceeding 10% and stemming from systematic errors.

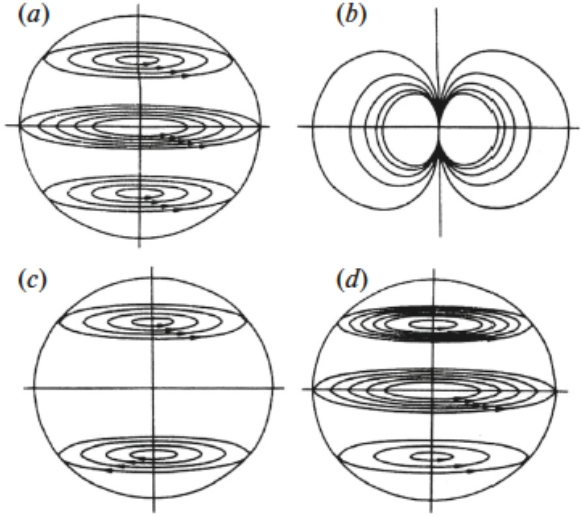

**Fig. 16.** (a) Initial toroidal field, (b) initial dipole field, (c) generated toroidal field, and (d) resulting toroidal field [50].

Such a correlation can readily be explained within the model of a magnetorotational explosion [47]. The explosion asymmetry leading to the kick velocity of a neutron star along the rotation axis may be due to the possible violation of mirror symmetry of the magnetic field in a differentially rotating star and the asymmetry of the neutrino flux. In [50], it is shown that, if the poloidal and toroidal components of the star magnetic field have different symmetry properties such that one of them is mirror-symmetric, while the other is mirror-antisymmetric, then the additional toroidal field induced by differential rotation enhances the toroidal component on one side of the equatorial plane and diminishes it on the other side (see Fig. 16; see also [51]).

In the process of field twisting upon collapse, the toroidal component may exceed the critical field strength $B_c$ at which the energy of electrons in the Larmor orbit is equal to the rest energy:

$$B_c = \frac{m_e^2 c^3}{e\hbar} = 4.4 \times 10^{13} \text{ G}. \quad (4)$$

For $B > B_c$, the probabilities for reactions induced by weak interaction become dependent on the magnetic field strength. In [52], the dependences in question for neutron beta decay were found to be

$$W_n = W_0[1 + 0.17(B/B_c)^2 + ...]$$
$$\text{for } B \ll B_c,$$
$$W_n = 0.77 W_0 (B/B_c) \quad \text{for } B \gg B_c. \quad (5)$$

In [53], the identical dependence on the magnetic field was used to estimate the neutrino mean free path and the neutrino opacity in a magnetic field. The magnetic-field dependence of the weak-interaction cross sections leads to the asymmetry of the neutrino flow and to the formation of fast flying pulsars because of the kick effect. Since the emergence of the asymmetry is due to the enhancement of the toroidal field around the rotation axis, the asymmetry of the neutrino flow is axially symmetric with respect to rotation axis, so that, owing to the kick effect, the pulsar acquires a velocity along the rotation axis as well. In [53], the pulsar velocity was estimated under the assumption that MDRI did not develop. By employing the magnetic-field dependence of the weak interaction cross sections in the form (5), we will now calculate the neutrino flux $H_\nu$ in the approximation of the neutrino thermal conductivity [54]. Specifically, we have

$$H_\nu = -\frac{7}{8} \frac{4acT^3}{3} l_T \frac{\partial T}{\partial r}, \quad (6)$$

where $l_T$ is the neutrino mean free path, which determines the neutrino opacity $\kappa_\nu$ as

$$\kappa_\nu = 1/(l_T \rho). \quad (7)$$

The anisotropy of the neutrino flux is determined as

$$\delta_L = \frac{L_+ - L_-}{L_+ + L_-}, \quad (8)$$

where $L_+$ and $L_-$ are the neutrino luminosities in two opposite directions along the axis of rotation. For power-law radial dependences of the temperature $T$ and $l_T$, the anisotropy of the neutrino flow and the acquired velocity of the neutron star can be calculated analytically in the form [53]

$$v_{nf} = \frac{2}{\pi} \frac{L_\nu}{M_n c} \frac{P B_{\phi 0}}{|B_p|} \left( 0.5 + \ln \left( \frac{20[s]}{P} \frac{|B_p|}{B_{\phi 0}} \right) \right), \quad (9)$$

$$L_\nu = \frac{0.1 M_n c^2}{20[s]}, \quad (10)$$

where $M_n$ is the neutron-star mass.

For $P = 10^{-3}$s and $L_\nu$ from (10), we find from (9) that

$$v_{nf} = \frac{2}{\pi} \frac{c}{10} \frac{P}{20[s]} x \ln \left( \frac{20[s]}{P} \frac{1}{x} \right)$$

$$\approx 1 \left[ \frac{km}{s} \right] x \ln \frac{2 \times 10^4}{x}, \quad (11)$$

where $x = B_{\phi 0}/|B_p|$ changes between 20 and $10^3$. The velocity acquired by the neutron star, $v_{nf}$, then ranges between 140 and 3000 km/s. This may explain the origin of the fastest pulsars.

## 6. POSSIBLE ORIGIN OF TWO NEUTRINO BURSTS FROM SN 1987A

Two pulses were detected in neutrino observations of SN 1987A. The first of them was recorded by the neutrino-radiation detector under Mont Blanc on February 23, 1987, at 2 : 52 : 36 UT [55, 56]. Five pulses of energy 7 to 11 MeV were detected within seven seconds.

The second pulse was observed after approximately 4.5 h at three facilities simultaneously. On February 23, 1987, a neutrino signal was recorded at 7 : 35 : 35 UT (°æ1 min) by the Kamiokande II underwater detector in Japan. The duration of the signal was 13 s, within which 11 electron events were detected in the energy range between 7.5 and 36 MeV [57]. The Irvine–Michigan–Brookhaven (IMB) neutrino telescope designed with the aim of searches for proton-decay reactions is situated in a Fairport salt mine, Ohio (USA). The observed signal consisted of eight neutrino events detected within six seconds and characterized by energies between 20 and 40 MeV. Concurrently, nine events were detected within the broader time interval between 7 : 35 : 40 and 7 : 35 : 50 UT [58]. At the Baksan underground scintillator telescope of Institute for Nuclear Research (USSR Academy of Sciences), five neutrino pulses of energy 10 to 25 MeV were detected within nine seconds on February 23, 1987, at 7 : 36 : 11 UT [59].

In optics, SN 1987A manifested itself on February 23, 1987, at 9 UT [60] approximately in 6 h after the first and 1.5 h after the second neutrino pulse. Three models were proposed for explaining two neutrino pulses from SN 1987A. The first, proposed in [61], associated the first pulse with the formation of a hot neutron star upon collapse and the second pulse with the collapse of a neutron star after its cooling and the formation of a black hole. The time of neutron star cooling was estimated in [61] at about an hour, which is strongly exaggerated. The calculations performed in [10] revealed that the neutrino-induced cooling of the neutron star did not exceed several tens of seconds; therefore, this model is unable to explain a nearly five-hour time lag between the two neutrino bursts. The second explanation of the two neutrino signals relied on the occurrence of collapse followed by the formation of a rotating magnetized star within the magnetorotational supernova model [62]. The first neutrino signal could be associated with the formation of such a neutron star, whereas the second signal was explained by the neutron-star collapse leading to the formation of a black hole. The time lag between the two neutrino signals had nothing to do with fast cooling but was due to the angular-momentum loss accompanying the magnetorotational-explosion process and leading to a decrease in the limiting mass of pulses via a mechanism that involves the fragmentation of a collapsing star was considered in [63]. This mechanism relies on the assumption of a very quick presupernova rotation, which should lead to fragmentation. However, this fragmentation may be hindered by the involvement of the magnetic field in angular momentum transfer, but its role was disregarded [64].

Let us consider in more detail the mechanism associated with a magnetorotational explosion. It should be noted that a large time interval between the formation of a rotating neutron star and the second collapse to a black hole can be obtained only in the absence of MDRI at a rather small initial magnetic-field strength of $10_8$ G. For want of an exact criterion for the development of MDRI, it can be assumed that, at such low magnetic fields, this instability is suppressed because of an overly strong excess of the gas pressure, $P_g >>> P_B$, and that, for the development of MDRI, the presence of some critical value $P_B/P_g \ll 1$ above which its development begins is required in addition to a strong excess of the toroidal component of the magnetic field over its poloidal component. The criteria for the development of instabilities in a rotating magnetized medium have not been studied conclusively. Observations of x-ray sources lead to the conclusion that, in the case of fulfillment of the hydrodynamic-stability condition, the presence of turbulent accretion disks is necessary. Balbus and Hawley [39] assumed that turbulence is a consequence of the development of MRI belonging to the type considered in [37, 38]. Many years of optical observations of the AM Her binary system containing the Her X-1 x-ray pulsar revealed [65, 66] the presence of long-term periods (lasting for about ten years) within which there was no x-ray flow, possibly because of the absence of turbulence in the accretion disk within these periods. We now assume that, for the onset of the development of a MDRI, the magnetic field should become higher than about $10^{14}$ G and that the average time of one field-generating turn is about 0.003 s. Within a time of about one hour, the field strength will then grow to a value at which the development of MDRI begins, entailing the emergence of a large outward angular-momentum flux and the formation of a magnetohydrodynamic shock wave that leads to supernova explosion and the collapse of the supernova remnant to a black hole. The properties of the two neutrino pulses and other observations are compatible with this model. The absence of any compact source in the SN 1987A remnant suggests the presence of a black hole there rather than of a neutron star. The release of energy in the first neutrino pulse is greater than $E_{nue1} \sim 10^{53}$ erg, while the average neutrino energy is less than that in the second pulse, where $E_{nue2} \sim 5 \times 10^{52}$ erg. The reason for this may be that the collapse is slower in the case of quick rotation than in the free fall case. Concurrently, neutrino-induced cooling hinders the growth of the temperature and reduces the average energy of emitted neutrinos. A collapse that ends up in black-hole formation first leads to the emission of high-energy neutrinos, but the neutrino luminosity decreases fast in the vicinity of the horizon, so that the energy of the second pulse turns out to be lower than the energy of the first pulse.

## 7. CONCLUSIONS

(i) The efficiency of rotational-energy transformation into the explosion energy in the course of a magnetorotational explosion is about 10%, which is sufficient for explaining the explosion energy of corecollapse supernovae.

(ii) The development of MDRI reduces strongly the time of a magnetorotational explosion at small values of the initial magnetic-field strength. (iii) A strong chaotic magnetic field of about $10_{14}$ G arises within a neutron star produced upon a magnetorotational explosion.

(iv) Jet formation occurs in the case where the initial magnetic field has a dipole configuration, and this may have some bearing on the origin of gamma ray bursts. A predominant ejection of matter near the equatorial plane arises if the initial magnetic field has a quadrupole configuration.

(v) A two-step collapse leading to black-hole formation may explain the observation of two neutrino pulses from SN 1987A. A magnetorotational explosion leads to a loss of the angular momentum, with the result that the newborn neutron star becomes unstable against a relativistic collapse leading to the formation of a black hole.

## ACKNOWLEDGMENTS

This work was supported in part by the Russian Foundation for Basic Research (project no. 17-02-00760), the Program for Support of LeadingScientific Schools (grant no. 6579.2016.2) and Program I.7P.